\documentclass[a4paper,oneside,english]{elsart}
\usepackage[T1]{fontenc}
\usepackage[latin1]{inputenc}
\usepackage{graphicx}

\makeatletter

\usepackage{babel}
\makeatother
\begin{document}

\begin{frontmatter}
\title{$Mn$ and $Fe$ impurities in $MgB_{2}$}

\author{P. Jiji Thomas Joseph and Prabhakar P. Singh}

\address{Department of Physics, Indian Institute of Technology  Bombay, Mumbai,
India 400076}

\ead{ppsingh@phy.iitb.ac.in}
\begin{abstract}
Based on first principles calculations, we show that $Mn$ impurities
are magnetic in $MgB_{2}$ due to  exchange-splitting of $d_{3z^2-1}$ band 
and they substantially modify 
 $B$ $p_{\sigma}$ and $p_{\pi}$ bands through  hybridization. 
Thus, $Mn$ impurities could act as strong magnetic scattering centers
leading to pair-breaking effects in $MgB_{2}$. In contrast, we find
$Fe$ impurities in $MgB_{2}$ to be nearly non-magnetic.
\end{abstract}
\begin{keyword}
impurity \sep magnetic moment \sep local-density approximation

\PACS 71.20.-b \sep 71.20.Be \sep 74.25.Jb \sep 71.15.Nc
\end{keyword}
\end{frontmatter}
Magnetic impurities in a conventional phonon-mediated single band
superconductor \cite{pr-108-1175} induce pair-breaking effects via
spin-flip scattering \cite{ag-1975}. However, for an unconventional
multi-band superconductor, pair-breaking seems to be most probable
from inter-band scattering \cite{prb-67-184515,prb-55-015146,prl-89-107002},
and thus it is not expected to depend solely on the exact atomic nature
of the impurity. Since the discovery of superconductivity in $MgB_{2}$ \cite{nature-410-63},
now confirmed to be a two-band superconductor \cite{prl-86-4374,prl-89-187003,prl-88-127002,prl-87-157002,prl-87-047001,prl-87-277001,prb-66-064511,ssc-121-479,nature-423-65},
previous studies of chemical substitutions have been largely confined
to examining the effects of inter-band and intra-band scattering of
the electrons. However, due to the orthogonality of $B$ $p_{\sigma}$
and $p_{\pi}$ orbitals, the inter-band scattering effects which could
lead to gap merging have not been observed in $MgB_{2}$ so far, although
the superconducting transition temperature $T_{C}$ showed a steady
decrease as a function of increasing impurity concentration \cite{prl-89-107002,prl-87-087005}.
Of particular interests were $Al$ \cite{nature-410-343,prb-65-020507,prb-68-094514,prb-71-174506,cmat-0507041}
and $3d$ transition-metal substitutions \cite{cmat-0104534,cmat-0510227,jpcm-14-12441,jpsj-70-1889,jap-93-8656,physC-370-39,sst-18-710,physC-424-79}
in the $Mg$ sub-lattice and $C$ substitutions \cite{prb-71-024533,pss-2-1656,prb-70-024504,physC-370-211,prb-71-134511}
in the $B$ sub-lattice of $MgB_{2}$. 

Of all impurity substitutions studied so far, $Mn$ shows the most
rapid decrease in $T_{C}$ as a function of its increasing concentration
in $MgB_{2}$ \cite{cmat-0104534,cmat-0510227,jpsj-70-1889}. However,
no evidence of inter-band scattering was observed, instead the role
of magnetic exchange splitting of the bands in $Mn$-substituted $MgB_{2}$
was emphasized \cite{cmat-0510227}. Earlier, it was shown that the
inter-band scattering would be important if the lattice was distorted
locally around the impurities \cite{prb-68-132505}. In this regard,
substituting $C$ impurities for $B$ atoms would have only a weak
effect on inter-band scattering because it does not change the local
symmetry. In contrast, substitutions in the $Mg$ plane could create
out-of-plane distortions and change the local point symmetry of nearby
$B$ atoms, resulting in a mixing of the in-plane $p_{x(y)}$ orbitals
and the out-of-plane $p_{z}$ orbital. Under these circumstances significant
$\sigma$-$\pi$ scattering would occur.

In this letter,  
we show that $Mn$ impurities
are magnetic in $MgB_{2}$ due to  exchange-splitting of $d_{3z^2-1}$ band 
and they substantially modify 
 $B$ $p_{\sigma}$ and $p_{\pi}$ bands through  hybridization. 
In contrast,
we find $Fe$ impurities in $MgB_{2}$ to be nearly non-magnetic.
Thus, the $Mn$ impurities could act as strong magnetic scattering
centers leading to pair-breaking effects in $MgB_{2}$. This may explain
the rapid decrease of $T_{c}$ in $Mg_{1-x}Mn_{x}B_{2}$ with increasing
$x$, while in $Mg_{1-x}Fe_{x}B_{2}$ a relatively slow decrease in
$T_{c}$ with $x$ may only be due to disorder-induced scattering
of the electrons. Before we discuss our results in detail, we briefly
describe the computational details. 

The normal metal electronic structure of the disordered alloys $Mg_{1-x}Mn_{x}B_{2}$
and $Mg_{1-x}Fe_{x}B_{2}$ are calculated using the Korringa-Kohn-Rostoker
(KKR) Green's function method formulated in the atomic sphere approximation
(ASA) \cite{turek}, which has been corrected by the use
of both the muffin-tin correction for the Madelung energy, needed
for obtaining an accurate description of ground state properties in
the ASA \cite{prl-55-600}, and the multipole moment correction to
the Madelung potential and energy which significantly improves the
accuracy by taking into consideration the non-spherical part of polarization
effects \cite{cmc-15-119,PRB-51-5773}. Chemical disorder was taken
into account by means of coherent-potential approximation (CPA) \cite{pr-156-809}.
The exchange and correlation were included within the local density
approximation (LDA) using the Perdew-Wang parameterization 
\cite{prb-45-13244} of the
many-body calculations of Ceperley and Alder \cite{prl-45-566}. During
the self-consistent procedure the reciprocal space integrals were
calculated by means of $1771$ $\mathbf{k}$-points in the irreducible
part of the hexagonal Brillouin zone (BZ), while the energy integrals
were evaluated on a semicircular contour in the complex energy plane
using $20$ energy points distributed in such a way that the sampling
near the Fermi energy was increased. When calculating the density
of states the number of $\mathbf{k}$-points were increased to $2299$.
The Green's function was evaluated for $1000$ points on a line in
the complex energy plane parallel to the real axis and was then analytically
continued towards the real axis. The atomic sphere radii of $Mg$
and $B$ were kept as $1.294$ and $0.747$, respectively of the Wigner-Seitz
radius. The sphere radii of the $3d$ substituents were kept the same
as that of $Mg$ itself. The overlap volume resulting from the blow
up of the muffin-tin spheres was approximately $10$\%, which is reasonable
within the accuracy of the approximation \cite{skriver-1984}. The
charge self-consistency iterations were accelerated using the Broyden's
mixing scheme \cite{prb-38-12807}, the CPA self consistency was speeded
up using the prescription of Abrikosov $et$ $al$ \cite{PRB-56-9319}.
The calculations for $Mg_{1-x}Mn_{x}B_{2}$ and $Mg_{1-x}Fe_{x}B_{2}$
alloys with $x=0.05$ were carried out at the experimental lattice
parameters of $MgB_{2}$ \cite{nature-410-63}, assuming a rigid underlying
lattice. The effects of local lattice relaxation as well as any possible
short-range ordering effects are not considered. Our results for $Mg_{1-x}Mn_{x}B_{2}$
and $Mg_{1-x}Fe_{x}B_{2}$ for $x=0.05$ are analyzed in terms of
total, sub-lattice resolved partial densities of states, and Bloch
spectral density $A_{B}(E,\mathbf{k})$ \cite{prb-21-3222} 
evaluated at high symmetry points of the hexagonal BZ. The electronic 
structure of body-centered cubic (bcc) $Fe$ is calculated with a similar 
setup using the experimental lattice constant $a=5.425$ a.u.. The density 
of states of bcc $Fe$ is calculated with $2470$ $\mathbf{k}$-points in 
the irreducible BZ. 

\begin{figure}
\includegraphics[%
  scale=0.3]{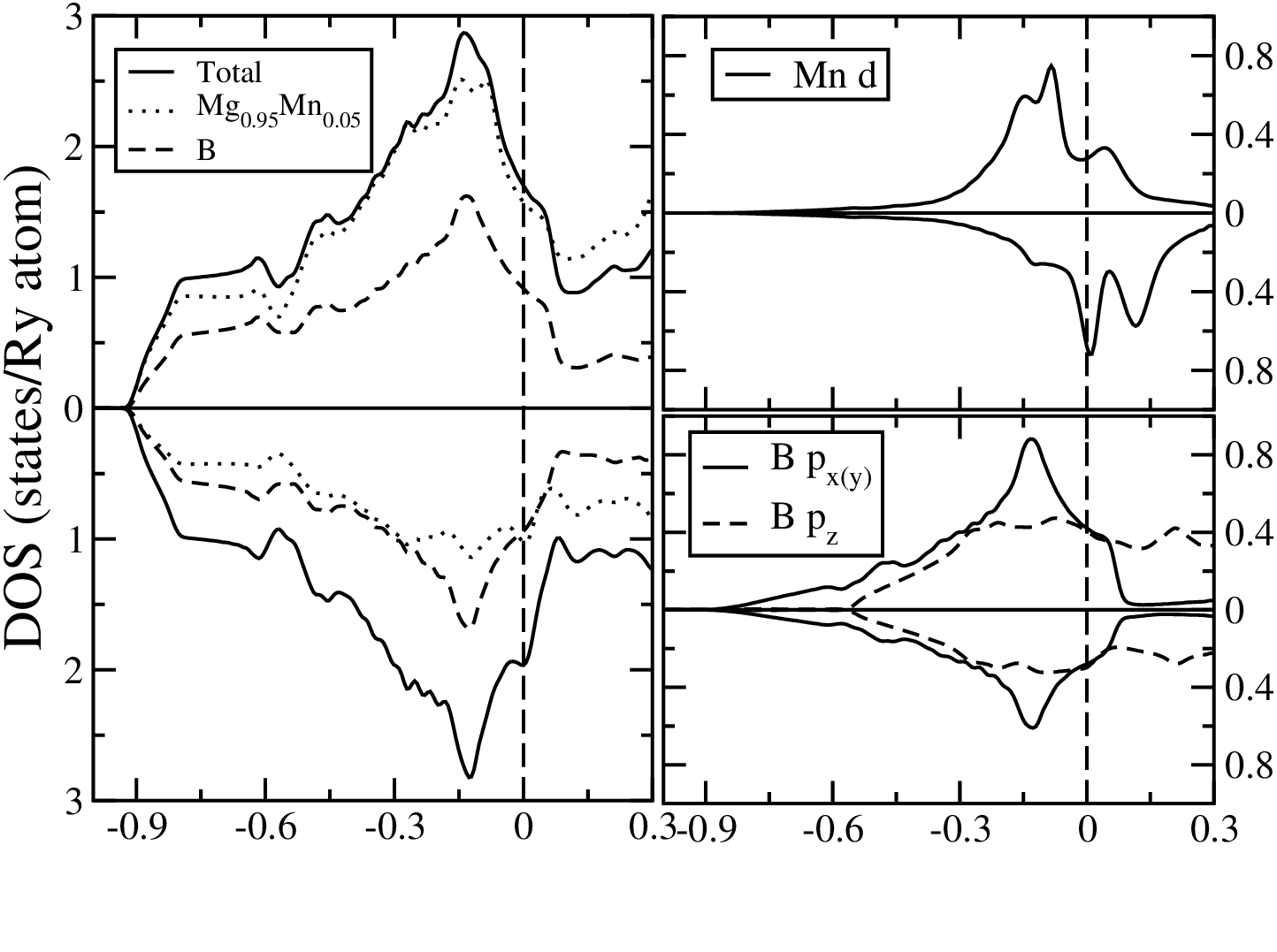}

\includegraphics[%
  scale=0.3]{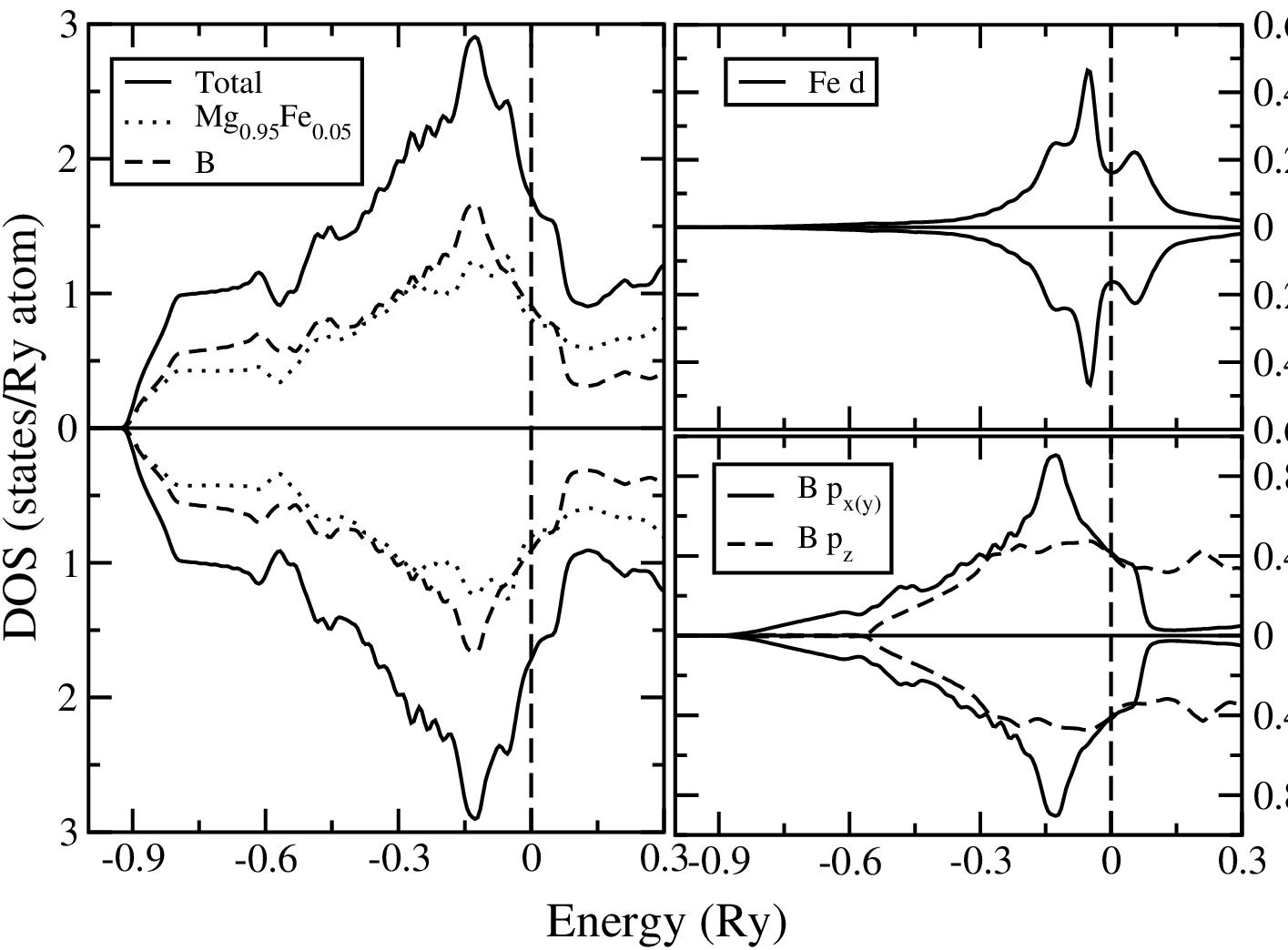}

\caption{\label{mnfe-tdos}The spin polarized, total and sub-lattice resolved
partial densities of states of $Mg_{0.95}Mn_{0.05}B_{2}$ (top panel)
and $Mg_{0.95}Fe_{0.05}B_{2}$ (bottom panel) alloys calculated as
described in the text. The transition-metal $d$ and $B$ $p$ densities
of states for the respective alloys are also shown in the side panels.
The dashed vertical line through energy zero represents the Fermi
energy.}
\end{figure}

The calculated density of states (DOS) at the Fermi energy, $E_{F}$,
of $MgB_{2}$ was determined to be $3.493$ states/Ry atom, which
is consistent with previous reports \cite{prl-86-4656,prl-87-087004}.
For $Mg_{0.95}Mn_{0.05}B_{2}$ and $Mg_{0.95}Fe_{0.05}B_{2}$ alloys,
the spin polarized calculations showed a lower total energy when compared
to the spin unpolarized calculations, suggesting the role of exchange
splitting of the states, in particular for $Mn$ substituted $MgB_{2}$.
The magnetic energies for $Mg_{0.95}Mn_{0.05}B_{2}$ and $Mg_{0.95}Fe_{0.05}B_{2}$
alloys are calculated to be $5.319$ mRy and $0.001$ mRy, respectively.
The corresponding local magnetic moment for $Mn$ is equal to $1.84$
$\mu_{B}$/atom, while $Fe$ tends to be feebly magnetic with a local
magnetic moment of $0.04$ $\mu_{B}$/atom. In Fig. \ref{mnfe-tdos}
we show the spin polarized, total and sub-lattice resolved partial
densities of states of $Mg_{0.95}Mn_{0.05}B_{2}$ and $Mg_{0.95}Fe_{0.05}B_{2}$
alloys. In the same figure we also show the transition-metal $d$
and $B$ $p_{x(y)}$ and $p_{z}$ DOS in these alloys. It is clear
from Fig. \ref{mnfe-tdos} that the exchange-split $Mn$ $d$ orbitals
in $Mg_{0.95}Mn_{0.05}B_{2}$ alloys make $Mn$ impurities magnetic
in $MgB_{2}$. We also find a substantial distortion in the majority
and minority $B$ $p_{x(y)}$ and $p_{z}$ bands due to their hybridization
with the $Mn$ $d$ orbitals. 
In addition, we find the $B$ $p$ band to have a very small exchange-splitting 
of $0.7$ mRy in $Mg_{0.95}Mn_{0.05}B_{2}$ but no splitting in  
$Mg_{0.95}Fe_{0.05}B_{2}$.

\begin{figure}
\includegraphics[%
  scale=0.3]{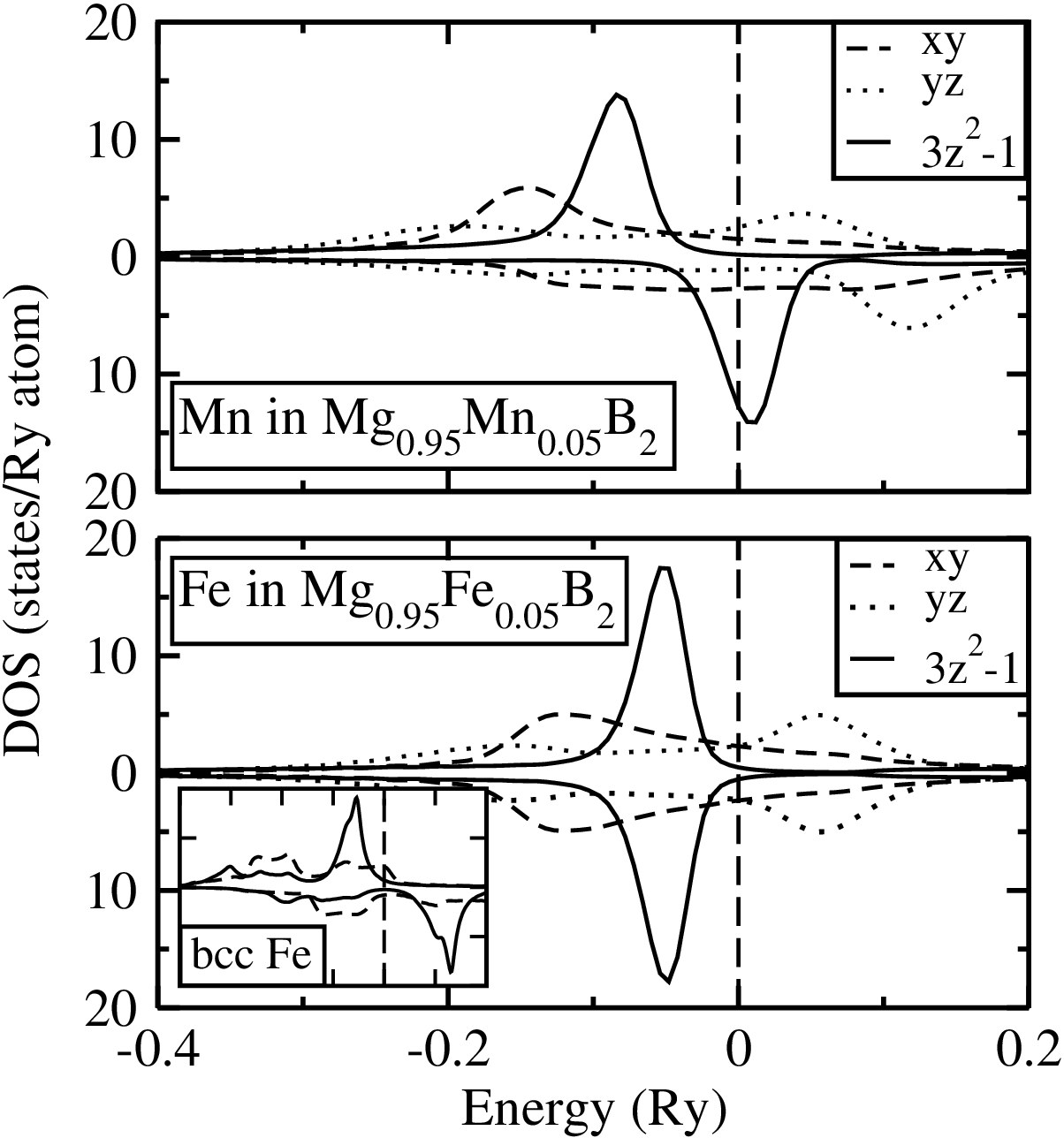}

\caption{\label{mnfe-ddos}The spin polarized, transition-metal $d$-projected
DOS in $Mg_{0.95}Mn_{0.05}B_{2}$ (top panel) and $Mg_{0.95}Fe_{0.05}B_{2}$
(bottom panel) alloys, resolved into three non-degenerate $xy$ $(x^{2}-y^{2})$,
$yz$ $(zx)$, and $3z^{2}-1$ symmetry components. 
The inset in the bottom panel shows the spin polarized, $d$-projected
DOS of bcc $Fe$, resolved into two non-degenerate $xy$ $(yz,zx)$ 
(dashed line) 
and $3z^{2}-1$ $(x^{2}-y^{2})$ (solid line) symmetry components. In the 
inset figure, the $x$-axis corresponds to energy ranging from $-0.4$ Ry 
to $0.2$ Ry, while the $y$-axis represents the DOS from $0$ to $12$ 
state/Ry/atom for both spins. 
The dashed vertical
line through energy zero represents the Fermi energy. Note that the
DOS in the figure corresponds to per transition-metal atom.}
\end{figure}

To gain further insight into such a contrasting behavior of $Mn$
and $Fe$ impurities in $MgB_{2}$, we show in Fig. \ref{mnfe-ddos}
the transition-metal $d$-projected DOS in $Mg_{0.95}Mn_{0.05}B_{2}$
and $Mg_{0.95}Fe_{0.05}B_{2}$ alloys, resolved into three non-degenerate
$xy$ $(x^{2}-y^{2})$, $yz$ $(zx)$, and $3z^{2}-1$ symmetry components.
We find that the band with $3z^{2}-1$ symmetry is the narrowest of
all the $d$ bands. Thus, a half-filled $3z^{2}-1$ band is more likely
to induce exchange splitting than the other bands. In the case of
$Mn$ the minority $3z^{2}-1$ band is nearly half-filled, resulting
in an exchange-split $d$ band. Our calculations show that the centers
of the $d\uparrow$ and the $d\downarrow$ bands are separated by
$0.114$ Ry leading to the Stoner exchange integral $I$ ($=\frac{d\uparrow-d\downarrow}{m}$)
to be $0.062$ Ry/$\mu_{B}$. However, in the case of $Fe$ the additional
one electron fills up the minority $3z^{2}-1$ band, as can be seen
from Fig. \ref{mnfe-ddos}, thereby making $Fe$ nearly non-magnetic
in $MgB_{2}$. In addition, following the prescription of Liechtenstein
$et$ $al$ \cite{jmmm-67-65}, we have calculated the on-site Heisenberg
exchange interaction $J_{0}$. The values of $J_{0}$ for $Mn$ and
$Fe$ alloys are found to be $0.7292$ mRy and $0.0001$ mRy respectively,
confirming $Mn$ to be magnetic and $Fe$ to be nearly non-magnetic
in $MgB_{2}$. 

It is instructive to compare the DOS of $Fe$ in 
$Mg_{0.95}Fe_{0.05}B_{2}$ with that of bcc $Fe$ as shown in the inset of 
Fig. \ref{mnfe-ddos}, where the $d$-projected DOS of bcc $Fe$ has been 
resolved into two non-degenerate $xy$ $(yz,zx)$ and 
$3z^{2}-1$ $(x^{2}-y^{2})$ symmetry components. The calculated 
magnetic moment of bcc $Fe$ is found to be $2.24$ $\mu_{B}$/atom. 
The most significant 
changes in the DOS of $Fe$ involve 
$x^{2}-y^{2}$ and $3z^{2}-1$ symmetry components. Due to the change in 
local symmetry, as one goes from bcc $Fe$ to hexagonal 
$Mg_{0.95}Fe_{0.05}B_{2}$, the narrow $x^{2}-y^{2}$ band of bcc $Fe$ 
hybridizes with the $B$ $p_{x(y)}$ bands of
$Mg_{0.95}Fe_{0.05}B_{2}$  and gets broadened as shown in 
Fig. \ref{mnfe-ddos}. The effects of hybridization on the $B$ $p_{x(y)}$ 
DOS can be seen in 
Fig. \ref{mnfe-tdos} in the form of a hump around 
$0.14$ Ry below the Fermi 
energy. On the other hand, the filling up of the minority $3z^{2}-1$ band 
by the extra electron results in an inward movement of the band by 
$0.15$ Ry, making magnetism unsustainable in 
$Mg_{0.95}Fe_{0.05}B_{2}$.

\begin{figure}[h]
~

~

\includegraphics[%
  scale=0.3]{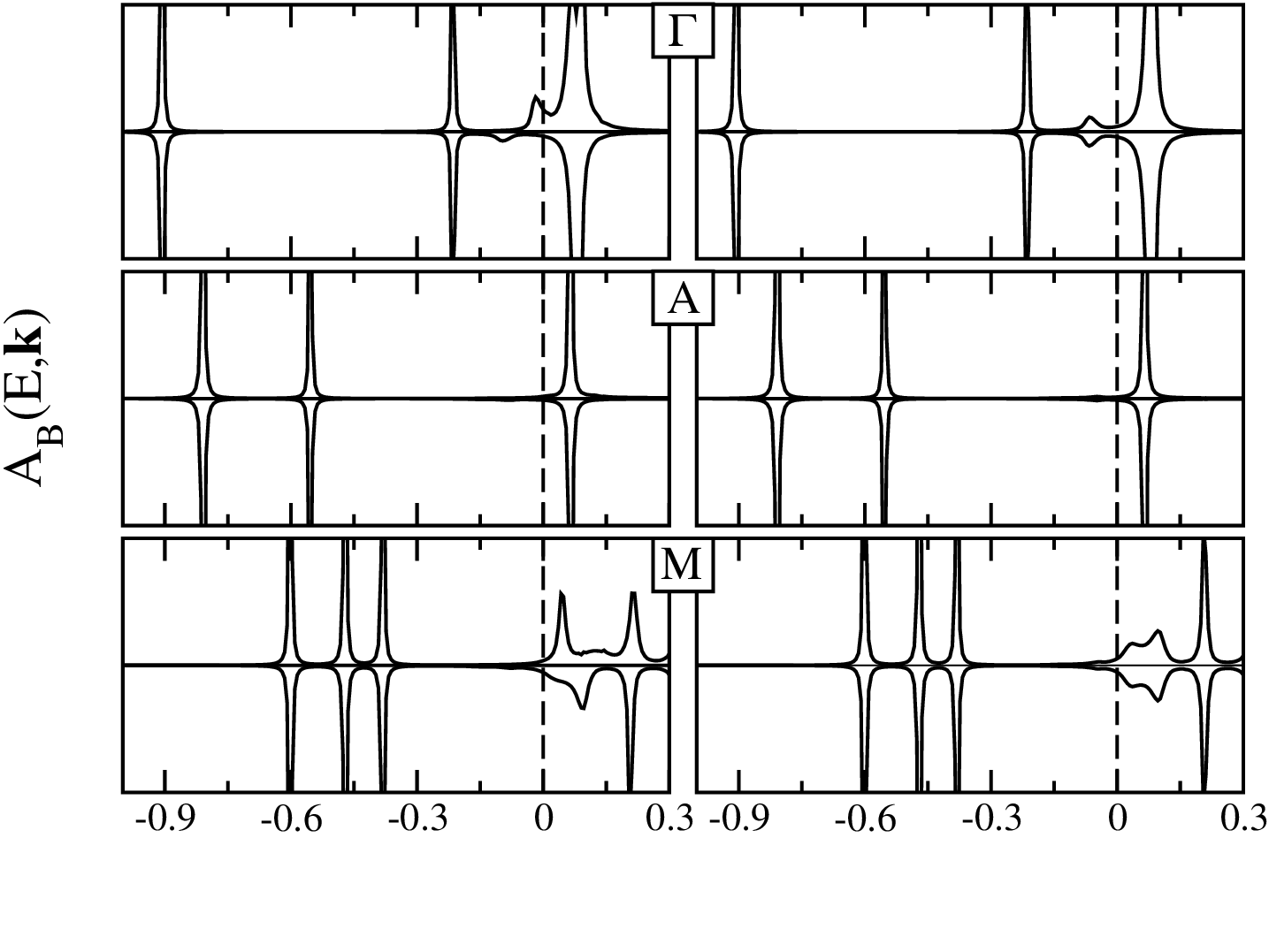}

\includegraphics[%
  scale=0.3]{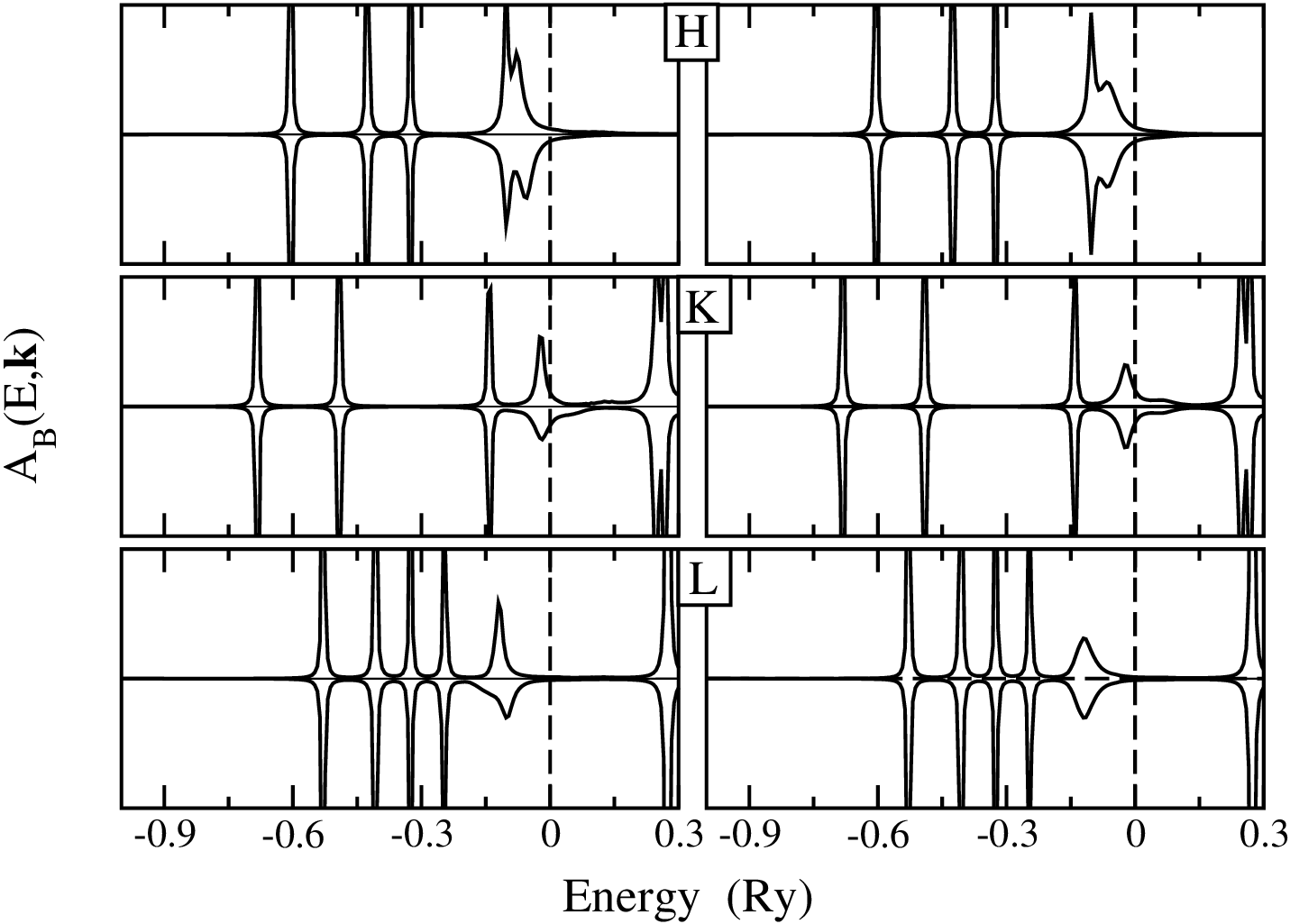}

\caption{\label{k-dos}The spin polarized, $\mathbf{k}$-resolved Bloch spectral
density $A_{B}(E,\mathbf{k})$ for $Mg_{0.95}Mn_{0.05}B_{2}$ (left) and
$Mg_{0.95}Fe_{0.05}B_{2}$ (right)  alloys at $\Gamma$, $A$ and $M$ (top
panel), and $H$, $K$, and $L$ points of the hexagonal Brillouin
zone. The dashed vertical line through energy zero represents the
Fermi energy.}
\end{figure}

Many of the superconducting properties of $MgB_{2}$ can be directly
related to its normal state electronic structure. In particular, an
understanding of how the additions of $Mn$ and $Fe$ impurities affect
the electronic states over the entire BZ can be very useful in this
regard. In Fig. \ref{k-dos} we show the spin polarized, $\mathbf{k}$-resolved
Bloch spectral density $A_{B}(E,\mathbf{k})$ for $Mg_{0.95}Mn_{0.05}B_{2}$
and $Mg_{0.95}Fe_{0.05}B_{2}$ alloys at $\Gamma$, $A$, $M$, $H$,
$K$, and $L$ points of the hexagonal BZ. As expected, the differences
between majority and minority $A_{B}(E,\mathbf{k})$ are much more
for $Mg_{0.95}Mn_{0.05}B_{2}$ alloys than for $Mg_{0.95}Fe_{0.05}B_{2}$
alloys over the entire BZ. New states, predominantly $d$ in character
are seen to form just below $E_{F}$ at $\Gamma$ and $K$ points.
The disorder-induced broadening of the doubly degenerate $B$ $p_{\sigma}$
states located between $-0.05<E<+0.05$ at $\Gamma$ and $B$ $p_{\pi}$
states just above $E_{F}$ at $M$ point can be seen clearly. At $M$
point, the lowest lying peak is characteristic of the $s$ band, while
the next two peaks correspond to the $p_{\sigma}$ bands. We find
that the $p_{\sigma}$ bands at $M$ point in ordered $MgB_{2}$ are
relatively far apart in comparison to the $Mn$ and $Fe$ substituted
systems. A change in the dispersion of bands is expected to affect
the three dimensionality of the Fermi surface \cite{prl-86-4656}
of the substituted materials (a detailed comparison will be published
elsewhere). We also note that in transition-metal diborides, such
as $TaB_{2}$ and others, a strong metal $d$ hybridization with the
$B$ $p$ bands is observed in $A$-$L$-$H$ plane of the BZ \cite{prb-64-174504,prb-64-144516}.
However, in the case of $Mn$ and $Fe$ impurities, the $d$ bands
are narrower and only the states that are close to $E_{F}$ are affected. 

In summary, we have carried out density-functional-based electronic
structure calculations to study the effects of $Mn$ and $Fe$ substitutions
on the $\sigma$ and $\pi$ bands of $MgB_{2}$. The self-consistent
calculations for $Mg_{0.95}Mn_{0.05}B_{2}$ and $Mg_{0.95}Fe_{0.05}B_{2}$
alloys show that $Mn$ forms a local magnetic moment of $1.84$ $\mu_{B}$/atom
while $Fe$ tends to remain feebly magnetic with a local magnetic
moment of $0.04$ $\mu_{B}$/atom. Further, the $\mathbf{k}$-resolved
Bloch spectral density of states at $\Gamma$, $A$, $M$, $H$, $K$,
and $L$ points of the hexagonal Brillouin zone show that for $Mn$
substituted alloys the $\sigma$ bands of $MgB_{2}$ are strongly
hybridized with the $Mn$ $d$ bands and undergo considerable spin-dependent 
modifications  
when compared to its $Fe$ counterparts. Given that $p_{\sigma}$
and $p_{\pi}$ orbitals are responsible for superconductivity in $MgB_{2}$,
$Mn$ impurities could, therefore, act as strong magnetic scattering
centers leading to pair-breaking effects. In contrast, the $Fe$ impurities
in $MgB_{2}$ are nearly non-magnetic with essentially no pair-breaking
effects.

One of us (PJTJ) would like to thank Dr. Andrei Ruban for providing
the KKR-ASA code. Discussions with Dr. Igor Mazin on the electronic
structure properties of $MgB_{2}$ is gratefully acknowledged.

\end{document}